\documentclass[conference]{IEEEtran}

\usepackage{cite}      

\usepackage[usenames,dvipsnames,svgnames,table]{xcolor}
\usepackage{graphicx}  

\usepackage{psfrag}    

\usepackage{url}       
\usepackage{tikz}

\usepackage{booktabs}

\usepackage{algorithm}
\usepackage{algpseudocode}

\usepackage{amsmath}   
\newcommand{\shorten}[1]{}
\newtheorem{proposition}{Proposition}

\newtheorem{definition}{Definition}

\newtheorem{example}{Example}
\newtheorem{observation}{Observation}

\begin{document}

\title{Minimal Header Overhead for Random Linear Network Coding}

%
\author{\authorblockN{
Danilo Gligoroski, Katina Kralevska, Harald {\O}verby}
\authorblockA{Department of Telematics, Faculty of Information Technology, Mathematics and Electrical Engineering, \\ Norwegian University of Science and Technology, Trondheim, Norway,\\ Email:
danilog@item.ntnu.no, katinak@item.ntnu.no, haraldov@item.ntnu.no}
}


\maketitle

\begin{abstract}
The energy used to transmit a single bit of data between the devices in wireless networks is equal to the energy for performing hundreds of instructions in those devices. Thus the reduction of the data necessary to transmit, while keeping the same functionality of the employed algorithms is a formidable and challenging scientific task. We describe an algorithm called Small Set of Allowed Coefficients (SSAC) that produces the shortest header overhead in random linear network coding schemes compared with all other approaches reported in the literature. The header overhead length is 2 to 7 times shorter than the length achieved by related compression techniques. For example, SSAC algorithm compresses the length of the header overhead in a generation of 128 packets to 24 bits, while the closest best result achieved by an algorithm based on error correcting codes has a header overhead length of 84 bits in $GF(16)$ and 224 bits in $GF(256)$. We show that the header length in SSAC does not depend on the size of the finite field where the operations are performed, i.e., it just depends on the number of combined packets $m$.
\\


\end{abstract}


%
\IEEEpeerreviewmaketitle

\vspace{0.5cm}
\section{Introduction}
The main feature of network coding is enabling the intermediate nodes in a multi-hop network to perform coding \cite{Ahlswede2000}. All nodes in the network excluding the sink nodes perform random linear mappings (RLNC \cite{Ho2006}) from input packets into output packets over a Galois Field of size $q$, $GF(q)$. The coding operations that are done over a packet are recorded in the packet header as a vector of coefficients. In the multi-hop networks, the vector of coefficients is updated at each node that performs network coding. The sink nodes decode the data based on the coefficients in the packet header.

One of the main challenges of implementing network coding is the header overhead imposed by the coding coefficients. When a source wants to send a large file, the file is split into several generations each consisting of $n$ packets. The length of the vector of coefficients is $n \log_2 q$ bits under RLNC in the finite Galois Field $GF(q)$. As the number of packets in a generation or the size of the Galois Field increases, the length of the header overhead due to the coding coefficients becomes significant. This affects the goodput of the system and can be a significant contribution to the system load for some network scenarios. 

Additionally, it is a known scientific fact that the energy used to transmit a single bit of data between the devices in ad hoc sensor networks is equal to the energy for performing 800 instructions in the devices \cite{Madden2002}. Thus the reduction of the data necessary to transmit, while keeping the same functionality of the employed algorithms is a formidable and challenging scientific task that implies that many applications will benefit by performing local computations rather than sending more bits. 

In this paper, we present a novel approach for practical network coding called Small Set of Allowed Coefficients (SSAC). The main contribution of this approach is that it generates the shortest compressed network coding header compared to related approaches reported in the literature. In SSAC the header length does not depend on the size of the finite field where the operations are performed, but only on the number of combined packets $m$.

The paper is organized as follows: Related work is presented in Section \ref{RelatedWork}. In Section \ref{Algorithm}, we present the Small Set of Allowed Coefficients algorithm. Experimental results are reported in Section \ref{Results}. Conclusions and future work are summarized in Section \ref{Conclusions}.

\section{Related Work} \label{RelatedWork}
Several papers in recent literature have addressed the problem of reducing the header overhead in network coding.

The first suggested solution was done in \cite{journals/tit/KoetterK08}. This approach finds a smaller vector subspace of the original vector space, and the coding is done just in that vector subspace. By this method, finding a proper subspace can be a computational challenge and decoding at a sink node is also a challenging task since every combination of source data should result in a distinct union subspace.

The concept of sparse coding is well known and it was first proposed in \cite{conf/isit/SiavoshaniKFA09} for header compression in network coding, to reduce the number of combined packets in one coded packet from $n$ to $m$ where $m < n$. This scheme uses parity-check matrices of error correcting codes to compress the header length down to $\mathcal{O}( m \log_2 n \log_2 q)$ bits. As noted in \cite{conf/isit/SiavoshaniKFA09}, the number of sources in sensor networks is large and a typical frame length is 30 bytes for data transmission. Consider a sensor network where 60 nodes send data. If RLNC in $GF(16)$ is performed, then 30 bytes per packet are used for recording the coding coefficients, i.e., the length of the header overhead is equal to the length of the useful data. Therefore, the authors of \cite{conf/isit/SiavoshaniKFA09} introduce the idea of compressing the coding vectors.
The length of the coding vector is reduced by limiting the number of packets that are combined in a coded packet denoted by $m$. However, limiting the number of packets being combined affects the invertibility of the matrix or decreases the probability of a redundant packet being innovative \cite{Blomer97therank,conf/icc/HeidePFM11,6892129,Pakzad05codingschemes}.

The authors in \cite{journals/icl/LiR10} proposed improved schemes for compression of the coding vectors by using erasure decoding and list decoding. The compressed header length under the erasure decoding scheme is $m+n/\log_2 q$. The header length becomes arbitrarily close to $m+O(\log_2 n)/\log_2 q$ when the list decoding scheme is used.
The both schemes are valid for moderate or large value of $m$.

Another completely different approach with a fixed and small header overhead was proposed in \cite{conf/vtc/ChaoCW10}. There, the header overhead is the seed for generating the coding coefficients with a known pseudo-random number generator (PRNG). This effectively reduces the header overhead to the size of the seed. However, as noted in \cite{conf/infocom/LiuWLZ10} this approach does not support re-encoding which is the crucial constituent of the random linear network coding.

A similar approach from the point of view of the extremity of reducing the overhead just to one symbol is proposed in \cite{journals/icl/ThomosF12}. There, the generation of the coding coefficients is based on modified Vandermonde matrices which can be determined by one symbol. However, the two big constraints of this design are: the network coding nodes should only perform addition operations and the generation size is upper bounded by $\log_2 q$ due to the cyclic property of the matrices.


We evaluate the presented approaches by using two metrics: the header length and the number of packets combined in a coded packet. An overview is given in Table \ref{SporedbaSite}. The features of SSAC are also presented in Table \ref{SporedbaSite} and we discuss them in the next Sections. Some of the presented methods do not support re-encoding or are valid for restricted set of $m$. Therefore, we compare SSAC with traditional RLNC and error correcting codes in Section \ref{Results}. 


\begin{table*}[t]
\caption{Comparison of header length in bits for different network coding schemes when the generation size is $n$}
\centering
\begin{tabular}{*{15}{c}}
\hline 
Scheme & Header length & Packets combined $m$ & \multicolumn{3}{c}{Operations in}\\
& & & Sources & Intemediate nodes & Destinations \\
\hline
RLNC & $n\log_2 q$ & $n$ & $GF(q)$ & $GF(q)$ & Gaussian elimination\\
\hline
Error correcting codes \cite{conf/isit/SiavoshaniKFA09} & $\mathcal{O}(m \log_2 n \log_2 q)$ & $ \log_2 n < m \leq \lfloor (n-k)/2 \rfloor$ & $GF(q)$ & $GF(q)$ & Berlekamp-Massey\\
\hline
Seed with PRNG \cite{conf/vtc/ChaoCW10} & Size of the seed & $n$ & $GF(q)$ & Do not support & Gaussian elimination\\
\hline
Erasure decoding \cite{journals/icl/LiR10} & $m \log_2 q + n$ & Moderate or large $m$ & $GF(q)$ & $GF(q)$  & Berlekamp-Massey\\
\hline
List decoding \cite{journals/icl/LiR10} & $m \log_2 q +\mathcal{O}(\log_2 n)$ & Moderate or large $m$ & $GF(q)$, $q$ is large & $GF(q)$, $q$ is large & Berlekamp-Massey\\
\hline
Vandermonde matrices \cite{journals/icl/ThomosF12} & $\log_2 q$ & $m \leq \log_2 q$ & $GF(q)$ & $GF(2)$ & Gaussian elimination\\
\hline
SSAC  & $m (\log_2 |Q| + \log_2 n)$ & $m$ & $GF(q)$ & $GF(q)$ & Gaussian elimination \\
\hline
\end{tabular}
\label{SporedbaSite}
\end{table*}

\section{The Algorithm: \emph{Small Set of Allowed Coefficients (SSAC)}} \label{Algorithm}
We denote by $GF(q)$ a finite field (Galois Field) with $q$ elements where $q$ is power of 2. 
It is known that for any finite field $GF(q)$ the set of all nonzero elements $GF(q)^\times = GF(q) \setminus {0}$ form a multiplicative cyclic group $(GF(q)^\times, \times)$. That means that any nonzero element $\beta \in GF(q)^\times$ can be represented as a power of a single element $\alpha \in GF(q)^\times$, i.e., $\beta = \alpha ^r$ for some $r \leq q$. Such a generator $\alpha$ is called a \emph{primitive element} of the finite field.

\shorten{
Let $(G,*)$ be a finite group with $q$ elements and operation $*$. It is known that for any subgroup $(P, *)$ where $P \subset G$ and $P$ has $p$ elements, $p$ must divide $q$. If $p$ is a prime number, the subgroup $P$ is called a Sylow-$p$ subgroup.
}

We consider that one or several sources send $n$ original data packets through a network where the source(s) and intermediate nodes can perform random linear network coding. We describe an algorithm which aims to provide a minimal header overhead for random linear network coding. The algorithm is based on utilizing a small set $Q \subset GF(q)$ of coefficients that multiply the original data. We formalize this with the following definition:

\begin{definition}
For a subset $Q$ of $GF(q)$ we say that it is a \emph{Small Set of Allowed Coefficients (SSAC)} if all operations of multiplication of the original data packets in the network coding procedures are performed \emph{only} by the elements of $Q$.
\end{definition}

Note that due to trivial reasons of impossible representation of the packet transformations, the set $Q$ cannot have just one element. 

The relation between the compression techniques presented in \cite{conf/isit/SiavoshaniKFA09} and \cite{journals/icl/LiR10} and our approach can be described as follows:  for the set $Q$ in \cite{conf/isit/SiavoshaniKFA09} and \cite{journals/icl/LiR10} they use all non-zero elements from the finite field $GF(q)$, while we use much smaller set. Namely, we use only two elements, i.e., $Q=\{q_0, q_1\}$ where both elements $q_0$ and $q_1$ are primitive elements in $GF(q)$. 

\shorten{
The second variant uses three element set $Q=\{q_{00}, q_{01}, q_{10}\}$ from the Sylow-$3$ subgroup of the multiplicative group $GF(q)^\times$. Note that the indexing of the elements in the three element set $Q$ is in binary notation $00$, $01$ and $10$ instead of $0$, $1$ and $2$. We use this binary indexing to denote which element multiplies the original data.
} 

Another crucial part of our method is the initial sparse encoding of the original data. Let us denote by $\mathbf{x}=(x_1, \ldots, x_n)$ a generation of $n$ original data packets. We set the level of sparsity to be a small number $m$, $m=2, 3$ or $4$ as reported in \cite{conf/isit/SiavoshaniKFA09}. In the beginning,  the source is generating a $k \times n$, ($k \geq n$) random sparse matrix
\begin{equation}
\label{eq:SparseInitialEncoding}
\mathbf{E}_{k \times n} = \left(
\begin{array}{c}
\mathbf{e}_1 \\
\vdots \\
\mathbf{e}_k \\
\end{array} \right)
\end{equation}
where every row-vector $\mathbf{e}_i$ is a sparse $n$-dimensional vector with just $m$ non-zero elements from the set $Q$. It uses $\mathbf{E}_{k \times n}$ to encode the initial data packets. However, in that encoding, due to the sparsity of the rows, instead of putting the whole rows as header overheads for each of the packets, it uses a special compression format.

\begin{definition}\label{Def:CSR}
We say that the row-vector $\mathbf{e}$ is encoded in \emph{Compressed Sparse Row (CSR)} format if it is presented in the form: $\mathbf{h}=(i_1 || j_1 || \ldots || i_m || j_m)$ where $i_\mu$ denotes the index of an element of the set $Q$ that is in the row vector $\mathbf{e}$ at the $j_\mu$ position, where $j_\mu$ is in binary format.
\end{definition}

Compressed Sparse Row (CSR) is applied frequently in mathematics and computing, and here we refer an interested reader to \cite{Barrett:94} as a good starting reference. Note that there is a slight difference between the described compressed formats in \cite{Barrett:94} and our format, since we apply the compressed sparse coding for each row separately (not for the whole matrix) due to the nature of the network coding paradigm where packets are transmitted through the network together with their header overhead. Thus we adopt the following convention:
\begin{definition}\label{Def:OurHeader}
All packets transmitted in the network have the format $\mathbf{h} || P$. The value of $\mathbf{h}$ is the header overhead as defined in Definition \ref{Def:CSR}, where the indices $j_\mu$ denote the indices of original data packets that are multiplied by the corresponding element $i_\mu \in Q$. The value of $P$ is the data payload for that packet.
\end{definition}

Without a proof we give here the following proposition:
\begin{proposition}
The length of $\mathbf{h}$ is $m (\log_2 |Q| + \log_2 n)$ bits.
\end{proposition}

It follows immediately that for a fixed value of $n$, the size of the header overhead $\mathbf{h}$ in bits is minimal if the size of the set $Q$ is minimal.

\begin{proposition}\label{Prop:MinHeaders}
If the set $Q$ has two elements, then for any number of original packets $n$, the size of the header overhead $\mathbf{h}$ encoded as in Definition \ref{Def:OurHeader} achieves the minimum value of $m (1+ \log_2 n)$ bits.
\end{proposition}

We give here a complete small example for coding of $n=8$ packets in $GF(16)$.

\begin{example}
 Let us use the the following irreducible polynomial $i(x) = x^4+x^3+1 $ in the finite field $GF(2^4)=GF(16)$. Let us choose the following two primitive elements: $q_0 = \mathbf{4} \equiv (0,1,0,0)_{16} \equiv 0 x^3 + 1 x^2 + 0 x + 0$ and $q_1 = \mathbf{14} \equiv (1,1,1,0)_{16} \equiv 1 x^3 + 1 x^2 + 1 x + 0$. Note that we denote the elements of the field with bold integers to distinguish them from ordinary integers. Note also that the integer binary representation in four bits corresponds with their polynomial representation with coefficients $\{0, 1\}$ and with polynomials of degree up to 3. 
 
 Let us consider that $n=8$ packets and $m=3$. If we have the following vector: $\mathbf{w} = (\mathbf{0},\mathbf{4},\mathbf{4},\mathbf{0},\mathbf{14},\mathbf{0},\mathbf{0},\mathbf{0} )$, then it is represented in the following CSR format: $\mathbf{h}=(0\, 001\, 0\, 010\, 1\, 100)$. 
Note that the spacing is just for readability, and that the indexing of the $n$ coordinates is done in the range from 0 to $n-1$.

Let us now suppose that a node has received $5$ random linear network coded packets, encoded with the following sparse row vectors: 
\begin{equation*}
\label{eq:SparseInitialEncodingExample}
\mathbf{E}_{5 \times 8} = 
\left(
\begin{array}{cccccccc}
 \mathbf{0} & \mathbf{0} & \mathbf{0} & \mathbf{4} & \mathbf{0} & \mathbf{0} & \mathbf{14} & \mathbf{4} \\
 \mathbf{4} & \mathbf{0} & \mathbf{4} & \mathbf{0} & \mathbf{0} & \mathbf{14} & \mathbf{0} & \mathbf{0} \\
 \mathbf{0} & \mathbf{0} & \mathbf{0} & \mathbf{14} & \mathbf{0} & \mathbf{14} & \mathbf{4} & \mathbf{0} \\
 \mathbf{0} & \mathbf{0} & \mathbf{4} & \mathbf{0} & \mathbf{14} & \mathbf{0} & \mathbf{14} & \mathbf{0} \\
 \mathbf{0} & \mathbf{14} & \mathbf{0} & \mathbf{14} & \mathbf{0} & \mathbf{0} & \mathbf{0} & \mathbf{14} \\
\end{array}
\right)
\end{equation*}

Note that $\mathbf{E}_{5 \times 8}$ is similar to  (\ref{eq:SparseInitialEncoding}), but since it is a node (not the source) the number of rows $k=5$ is less than $n$. We say that number $k$ in every node represents \emph{the number of buffered packets in that node}. If the node combines all of its buffered packets (in this case $5$ packets) it may find a new innovative packet which is a combination of the original data packets, with a vector that has only $m$ nonzero elements from the set $Q$. Indeed, if the buffered packets are combined with the following vector: $\mathbf{x}=(\mathbf{1},\mathbf{0},\mathbf{0},\mathbf{1},\mathbf{5})$ then the new innovative packet is $\mathbf{e}_6 = \mathbf{w} = \mathbf{x} \cdot \mathbf{E}_{5 \times 8} = (\mathbf{0},\mathbf{4},\mathbf{4},\mathbf{0},\mathbf{14},\mathbf{0},\mathbf{0},\mathbf{0} )$. By \emph{innovative} we mean that it is linearly independent from all existing rows in the matrix $\mathbf{E}_{5 \times 8}$. Then the node generates the compressed header $\mathbf{h}_6=(0\, 001\, 0\,  010\, 1\, 100)$ and with the vector $\mathbf{x}$ encodes the buffered $5$ packets producing the innovative data payload $P$. It sends the packet $\mathbf{h}_6 || P$ as presented in Definition \ref{Def:OurHeader} .
\end{example}

We systematize the previous example in a form of a precise step by step algorithm SSAC in Table \ref{Alg:MinHeaderOverheads}.
\begin{table}[!h]
\centering
\caption{An algorithm for network coding that generates an innovative packet $(\mathbf{h}_{k+1} || P_{k+1})$ where the header overhead $\mathbf{h}_{k+1}$ has a minimal length in bits}\label{Alg:MinHeaderOverheads}
\begin{tabular}{|c|}
  \hline
   \parbox{7.5cm}{\center \textbf{Algorithm: Small Set of Allowed Coefficients (SSAC)} \vspace{0.1cm}}\\ %
  \hline
  \parbox{7.5cm}{\vspace{0.1cm} {\bf Input:} $n$, $k$, $m$, $GF(q)$, $Q=\{q_0, q_1\}$, (or $Q=\{q_{00}, q_{01}, q_{10}\}$), \\ $\mathbf{Data}=\{(\mathbf{h}_1 || P_1), \ldots, (\mathbf{h}_k || P_k)\}$ \vspace{0.1cm}}\\
  \hline
  \parbox{7.5cm}{\vspace{0.1cm} {\bf Output:} $(\mathbf{h}_{k+1} || P_{k+1})$ \vspace{0.1cm}}\\
  \hline
  \begin{tabular}{l}
  \parbox{7.5cm}{\vspace{0.1cm} 1. Set $\mathbf{H} = (\mathbf{h}_1, \mathbf{h}_2, \ldots, \mathbf{h}_k)^T$; }\\
  \parbox{7.5cm}{\vspace{0.1cm} 2. Set $\mathbf{E} = (\mathbf{e}_1, \mathbf{e}_2, \ldots, \mathbf{e}_k)^T$; where $\mathbf{e}_i$ are CSR forms of $\mathbf{h}_i$}\\
  \parbox{7.5cm}{\vspace{0.1cm} 3. Set $\mathbf{P} = (P_1, P_2, \ldots, P_n)^T$; }\\
  \parbox{7.5cm}{\vspace{0.1cm} Repeat }\\
  \parbox{7.5cm}{\vspace{0.1cm} \ \ \ \ \ \ \ \ 4. Set $\mathbf{w} \leftarrow \mbox{RandomSparseVector}(m,n,Q)$; }\\
  \parbox{7.5cm}{\vspace{0.1cm} \ \ \ \ \ \ \ \ 5. Find $\mathbf{x}$ such that $\mathbf{w} = \mathbf{x} \cdot \mathbf{E}$; }\\
  \parbox{7.5cm}{\vspace{0.1cm} Until found $\mathbf{x}$; }\\
  \parbox{7.5cm}{\vspace{0.1cm} 6. Set $\mathbf{e}_{k+1} = \mathbf{w}$; }\\
  \parbox{7.5cm}{\vspace{0.1cm} 7. Set $\mathbf{h}_{k+1} = \mbox{CompressedSparseRow}(\mathbf{e}_{k+1})$; }\\
  \parbox{7.5cm}{\vspace{0.1cm} 8. Set $P_{k+1} = \mathbf{x} \cdot \mathbf{P}$; }\\
  \parbox{7.5cm}{\vspace{0.1cm} 9. Return: $(\mathbf{h}_{k+1} || P_{k+1})$. \vspace{0.1cm}}\\
  \end{tabular}\\
 \hline
  \end{tabular}
\end{table}

\subsection{Efficiency of SSAC}
The procedure called $\mbox{RandomSparseVector}(m,n,Q)$ in Step 4 returns one random $n$ dimensional vector that is sparse and has exactly $m$ nonzero coordinates from the set $Q$. It is important to notice that the algorithm is a probabilistic one, and in Step 5 it attempts to find a solution $\mathbf{x}$ of a linear matrix-vector equation 
\begin{equation}
\label{eq:MatVecSparse}
\mathbf{w} = \mathbf{x} \cdot \mathbf{E}.
\end{equation} Depending on the values of $k$, $m$, $n$ and $\mathbf{E}$ it is not always possible to find such a linear solution $\mathbf{x}$. One work that addresses the problem of existence of solutions of equation (\ref{eq:MatVecSparse}) is \cite{journals/corr/abs-1106-0365}. We summarize the findings in \cite{journals/corr/abs-1106-0365} adopted for our SSAC algorithm with the following Observation:
\begin{observation}
\label{thm:PacketsInBufferAndProbability}
Let $\mathbf{E}_{k \times n}$ be a random sparse matrix where the sparsity of each row is such that there are exactly $m$ nonzero elements from the set of coefficients $Q$. If the number of rows $k$ is 
\begin{equation}\label{eq:ApproxInBuffer}
k = k_{opt} \approx m \log \frac{n}{m},
\end{equation}
then there exists a sparse vector $\mathbf{w}$ (with a sparsity of having exactly $m$ nonzero elements from the set of coefficients $Q$) and a solution to the equation (\ref{eq:MatVecSparse}) with a probability $1-\epsilon$, where $\epsilon$ is a small value.
\end{observation}

We emphasize that our work is the first one that explicitly addresses the efficiency of the header overhead compression algorithm in every intermediate node in conjunction with the number of buffered packets $k$ in that node. In other related works such as \cite{journals/icl/LiR10} and \cite{conf/isit/SiavoshaniKFA09} the number of buffered packets in the intermediate nodes as a factor for the efficiency of the algorithm is not addressed at all.

Another important question is the efficiency of the SSAC algorithm for finding a solution of the equation (\ref{eq:MatVecSparse}). We experimentally measured the number of attempts in the Repeat-Until loop of the SSAC algorithm for the sparsity values $m=2, 3$ or $4$, the number of packets in the range from 16 to 128, and the number of packets in the nodes $k \approx m \log \frac{n}{m}$. The results are given in the next Section. 


\subsection{Probability of successful decoding at the destination}
Another important aspect in the analysis of SSAC is the probability of successful decoding of the original data. As reported in \cite{conf/icc/HeidePFM11}  and \cite{Pakzad05codingschemes} the probability of successful decoding depends on the sparsity $m$, the size of the finite field $q$ and the overhead $O$ (for $n$ encoded packets, the receiver needs $n+O$ packets in order to decode them successfully). 
Our experiments presented in the next Section confirm that SSAC behaves in the same way as reported in \cite{conf/icc/HeidePFM11,Pakzad05codingschemes,6892129}. The probability of successful decoding is lower when using sparse codes compared to dense codes. On the other hand, the probability of successful decoding increases with the overhead and the field size.

\section{Experimental Results} \label{Results}

In this Section we illustrate the performance of SSAC and compare it with RLNC and the algorithm based on error correcting codes.

Network coding is usually performed in $GF(16)$ and $GF(256)$, therefore we make experiments for these finite fields for different generation sizes. In order to check the correctness of the presented results, we give the concrete values of the finite fields and the primitive elements that we used. The irreducible polynomial for $GF(16)$ is $i(x) = x^4+x^3+1$. For a two-element Small Set of Allowed Coefficients we choose $Q=\{\textbf{4}, \textbf{14} \}$. 

The irreducible polynomial for $GF(256)$ is $i(x) = x^8+x^6+x^3+x^2+1$. For a two-element Small Set of Allowed Coefficients we choose $Q=\{\textbf{21}, \textbf{43} \}$. 

\begin{figure}[t]
\centering
\includegraphics[width=3.4in]{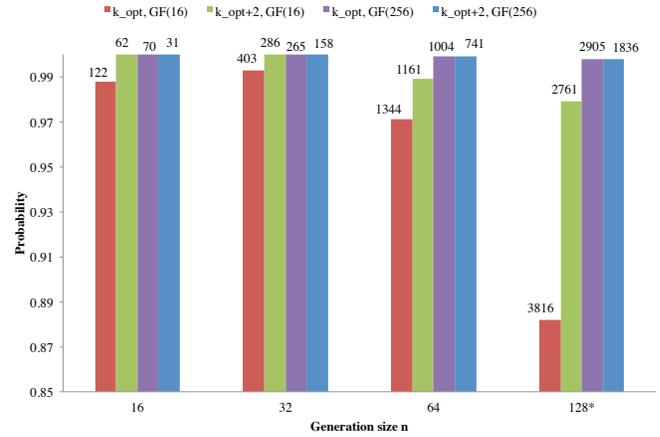}
\caption{Probability of solution existence and the number of attempts to find the solution when $m=2$ for different values of $n$ and $k_{opt}$ in $GF(16)$ and $GF(256)$.
Note that the height of the bars denotes the probability, but the values above the bars denote the average number of attempts in SSAC algorithm. Also note that for $n=128$ in our experiments we took $k=k_{opt} + 8$.
}
\label{attempts}
\end{figure}

First, we investigate the probability of solution existence for different values of $m$, $n$ and $k_{opt}$ in $GF(16)$ and $GF(256)$. Figure \ref{attempts} shows the probability of solution existence and the number of attempts in the Repeat-Until loop for the aforementioned parameters for $m=2$. We see that the probability of solution existence increases with $k_{opt}$ and the size of the field. As we can see that the number of attempts in the Repeat-Until loop varies in the range from 30 up to 3816.

\begin{figure}
\centering
\includegraphics[width=3.4in]{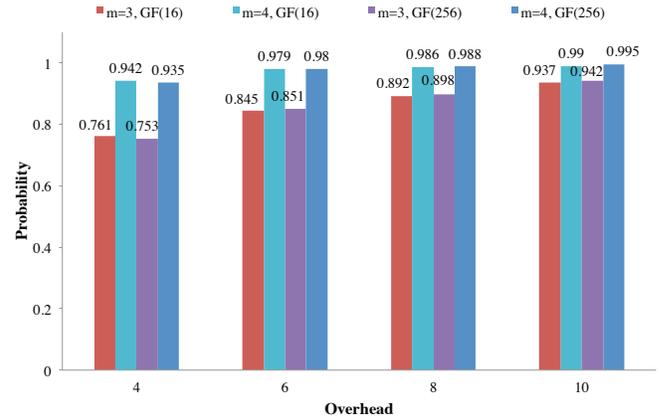}
\caption{Probabilty of a full rank matrix as a function of the overhead when $n$=16 for different $m$ in $GF(16)$ and $GF(256)$}
\label{ov34gf16256}
\end{figure}

Next we investigate the probability that the received matrix at the destination has a full rank  when it has more than $n$ rows, i.e., it has some overhead. Figure \ref{ov34gf16256} shows the probability of a successful decoding as a function of the overhead for $n=16$ in $GF(16)$ and $GF(256)$ when $m=3,4$ and the destination receives overhead of 4, 6, 8 or 10 extra packets. From Figure \ref{ov34gf16256} we can see that probability of having a successful decoding increases with $m$ and the overhead.

\begin{figure}
\centering
\includegraphics[width=3.4in]{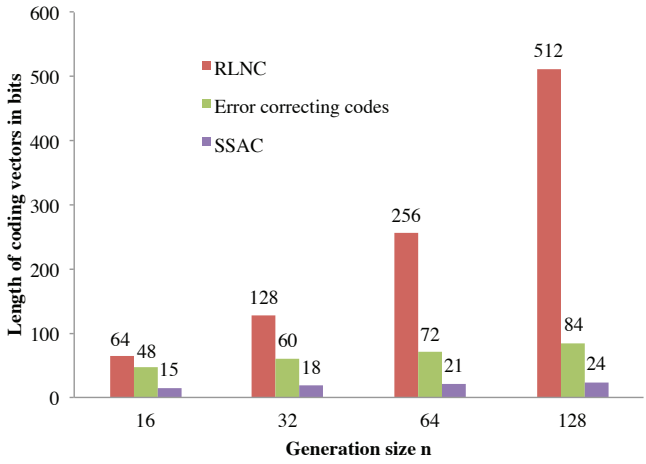}
\caption{Length of the compressed coding vectors in bits as a function of the number of packets in a generation $n$ when $m$=3 in $GF(16)$}
\label{m3gf16}
\end{figure}

\begin{figure}
\centering
\includegraphics[width=3.4in]{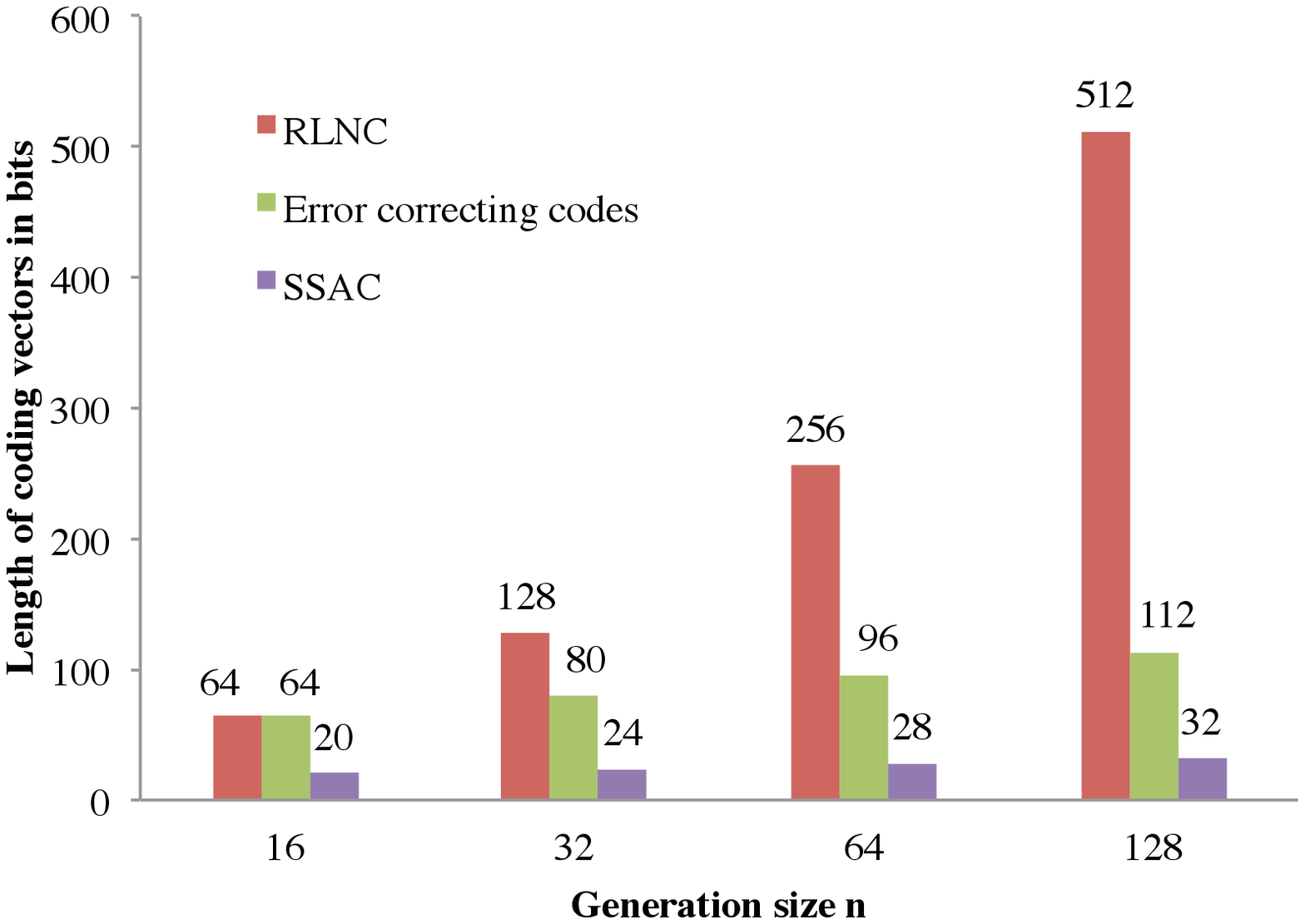}
\caption{Length of the compressed coding vectors in bits as a function of the number of packets in a generation $n$ when $m$=4 in $GF(16)$}
\label{m4gf16}
\end{figure}

\begin{figure}
\centering
\includegraphics[width=3.4in]{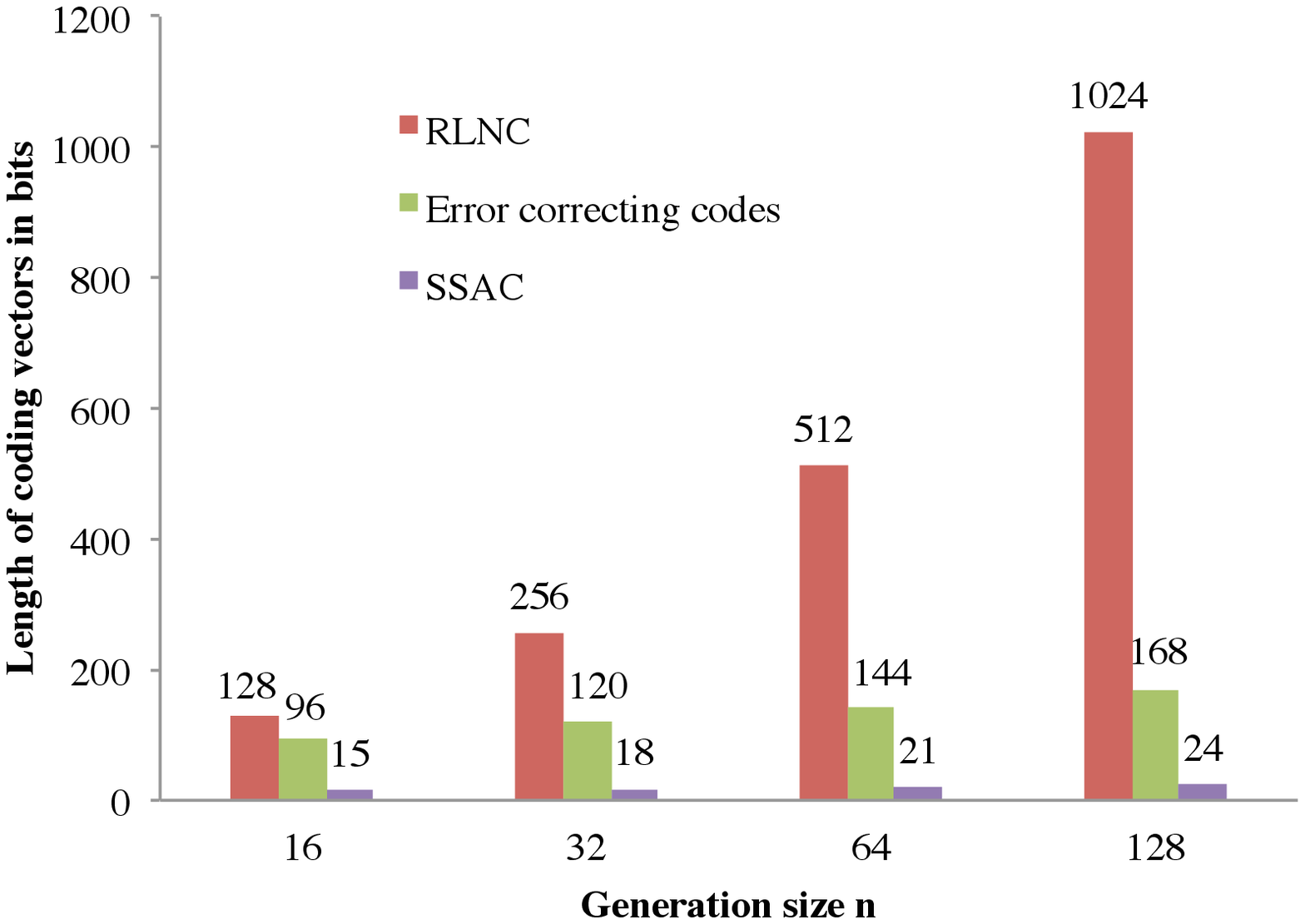}
\caption{Length of the compressed coding vectors in bits as a function of the number of packets in a generation $n$ when $m$=3 in $GF(256)$}
\label{m3gf256}
\end{figure}

\begin{figure}
\centering
\includegraphics[width=3.4in]{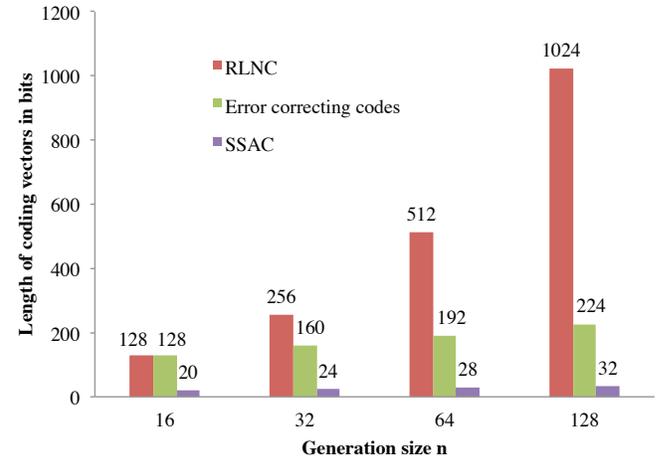}
\caption{Length of the compressed coding vectors in bits as a function of the number of packets in a generation $n$ when $m$=4 in $GF(256)$}
\label{m4gf256}
\end{figure}

\begin{figure}
\centering
\includegraphics[width=3.4in]{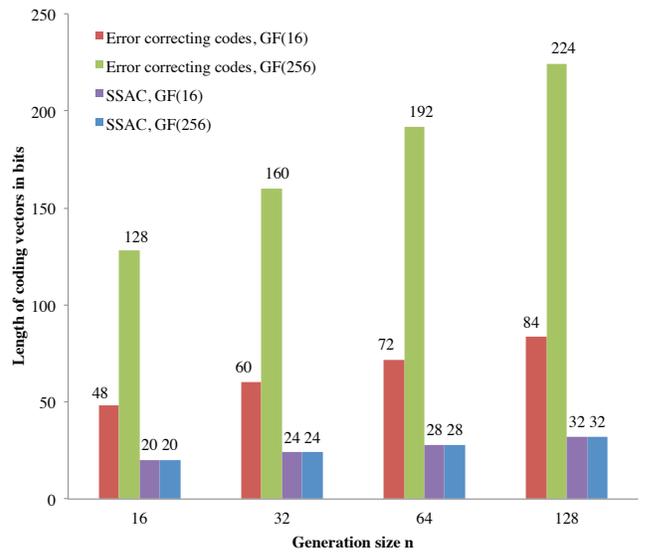}
\caption{Length of the compressed coding vectors in bits as a function of the number of packets in a generation $n$ for $m$=4 in $GF(16)$ and $GF(256)$}
\label{m4gf16_256}
\end{figure}

\begin{figure}
\centering
\includegraphics[width=3.4in]{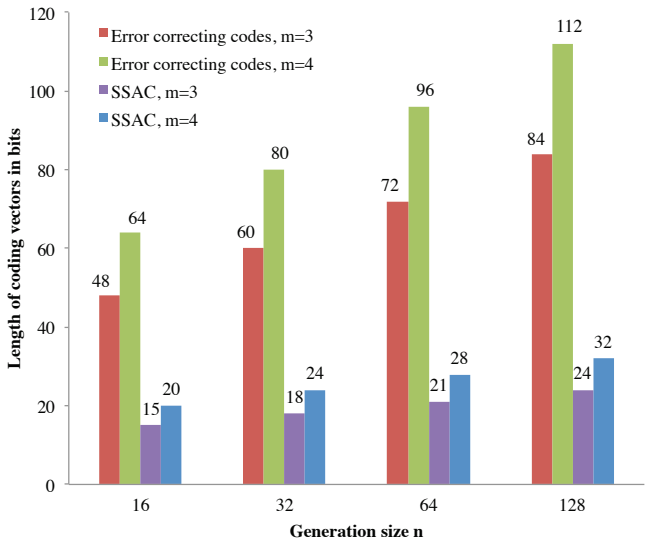}
\caption{Length of the compressed coding vectors in bits as a function of the number of packets in a generation $n$ for $m$=3 and $m$=4 in $GF(16)$}
\label{m34gf16}
\end{figure}

\begin{figure}
\centering
\includegraphics[width=3.4in]{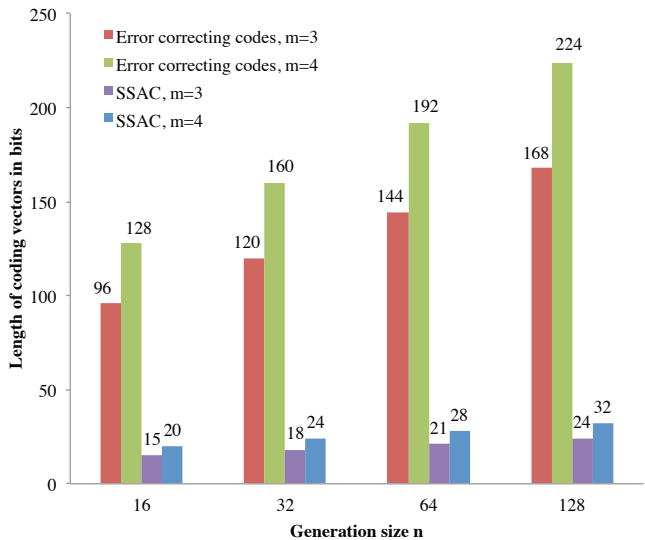}
\caption{Length of the compressed coding vectors in bits as a function of the number of packets in a generation $n$ for $m$=3 and $m$=4 in $GF(256)$}
\label{m34gf256}
\end{figure}

We also compare the length of the coding vectors in bits for different schemes. As we stated previously, we compare the performance of SSAC with the most relevant approaches, i.e., traditional RLNC and error correcting codes based approach. Figure \ref{m3gf16} to Figure \ref{m34gf256} depict the length of the coding vectors in bits versus the generation size $n$ for different $m$ and finite fields. The length of the header overhead increases with the generation size in all three approaches as shown in Figure \ref{m3gf16} to Figure \ref{m34gf256}. However, SSAC achieves around 42 times shorter header overhead compared to RLNC in $GF(256)$ as presented in Figure \ref{m3gf256}. Compared to the method with error correcting codes, SSAC produces around 7 times shorter header overheads when $m$=3 in $GF(256)$. Figure \ref{m3gf256} and Figure \ref{m4gf256} show that the ratio between the header lengths of SSAC and RLNC is decreasing as $m$ is increasing and the coding is performed in the same finite field. On the other hand, the ratio between the header lengths of SSAC and Error correcting codes approaches is the same for different $m$ in the same finite field.

Figure \ref{m4gf16_256} shows the length of the coding vectors in bits versus the generation size in $GF(16)$ and $GF(256)$ when $m$ is fixed to 4. It is well-known fact that the header overhead increases with the finite field. However, the finite field size does not have an impact on the header overhead when the coding coefficients are generated with SSAC as shown in Figure \ref{m4gf16_256}. For instance, the length of the coding vector is the same in $GF(16)$ and $GF(256)$ for the same $m$. This can be considered as a big advantage of the SSAC algorithm.

Figure \ref{m34gf16} and Figure \ref{m34gf256} show how the length of the header overhead depends on $m$. If $m$ increases, then the length of the header overhead increases as presented in Figure \ref{m34gf16} and Figure \ref{m34gf256}. The step of increasing of the header overhead length in dependence on $m$ is around 8 bits for the SSAC  algorithm. On the other hand, this step is significantly bigger for the algorithm based on error correcting codes.

\vspace{0.5cm}

\section{Conclusions} \label{Conclusions}
In this paper we proposed the algorithm Small Set of Coefficients (SSAC). The SSAC achieves the shortest header overhead in random linear network coding schemes compared to other approaches reported in the literature. We compared the SSAC with RLNC and Error correcting codes. The header overhead length with SSAC is 2 to 7 times shorter than the length achieved by these compression techniques. We show that the header length does not depend on the size of the finite field where the operations are performed, i.e., it just depends from the number of packets combined $m$.

As a future work we suggest the following: 1. To establish a precise empirical correlation between the parameters $m$, $n$, $k$, $q$ and the probability of finding solutions of the equation (\ref{eq:MatVecSparse}); 2. To improve the efficiency of the algorithm SSAC by replacing the randomized search approach in Step 4 with a faster direct algorithm for finding suitable $\mathbf{w}$ and $\mathbf{x}$ in equation (\ref{eq:MatVecSparse}); 3. To introduce and examine a realistic metric of energy consumption in the source and intermediate nodes for producing and transmitting the headers; and 4. To investigate the relation between the elements in the set $|Q|$ and the efficiency of the algorithm.

\bibliographystyle{plain}
\bibliography{refer}

\end{document}